\documentclass{article}
\usepackage{geometry}
\usepackage[utf8]{inputenc}
\usepackage{graphicx}
\usepackage{subfigure}
\usepackage{amssymb}
\usepackage{amsmath}
\usepackage{epstopdf}
\usepackage{cite}
\usepackage{bbm}
\usepackage{multirow}
\usepackage{enumerate}
\usepackage{xcolor}
\usepackage[hyperindex=true,
          pdfstartview=FitH,
          bookmarksnumbered=true,
          bookmarksopen=true,
          citecolor=blue,
          linkcolor=blue,
          colorlinks=true,
          pdfborder=001,
          unicode]{hyperref}

\newcommand{\ba}{\begin{align}}
\newcommand{\ea}{\end{align}}
\newcommand{\be}{\begin{equation}}
\newcommand{\en}{\end{equation}}
\newcommand{\bea}{\begin{eqnarray}}
\newcommand{\ena}{\end{eqnarray}}
\newcommand{\nn}{\nonumber}
\usepackage{graphicx} 

\title{Dynamics of null particles and shadow for general rotating black hole}
\author{Kun Meng\\School of Physics and Electronic Information,
	Weifang University, Weifang 261061, China \\Xi-Long Fan\footnote{xilong.fan@whu.edu.cn} \\School of Physics and Technology, Wuhan University, Wuhan, Hubei 430072, China \\Song Li \\Shanghai Astronomical Observatory, Shanghai, 200030, China; \\School of Astronomy and Space Science, University of Chinese Academy of Sciences,\\
	Beijing, 100049, China \\Wen-Biao Han \\School of Fundamental Physics and Mathematical Sciences, \\Hangzhou Institute for Advanced Study, UCAS, Hangzhou 310024, China;
	\\ Shanghai Astronomical Observatory, Shanghai, 200030, China;\\ School of Astronomy and Space Science, University of Chinese Academy of Sciences,\\ Beijing, 100049, China;\\ International Centre for Theoretical Physics Asia-Pacific, Beijing/Hangzhou, China\\ Hongsheng Zhang\footnote{sps\_zhanghs@ujn.edu.cn} \\School of Physics and Technology, University of Jinan, Jinan, China}

\date{\today}

\begin{document}

\maketitle

\begin{abstract}
The Johannsen black hole (BH) is a generic rotating BH admitting three constants of motions ( energy, angular momentum, and Carter constant) and is characterized by four deviation parameters besides mass and spin, which could be a model-independent probe of the no-hair theorem. We systematically study the dynamics of null particles around Johannsen BH, revealing the effects of the deviation parameters on the BH shadow as well as the effects of spin. By using the shadow boundaries of M87* and SgrA*, for the first time, the deviation parameters of those BHs are constrained. The detailed results depend on the spin $a$ and inclination angle $ \theta_0$. Assuming $a=0.2$ and $\theta_0=15^{\circ}$, the deviation parameter $\alpha_{13}$ are constained within $\sim $ [-3.5, 6] for M87* observation and [-3, 0.5] for SgrA* observation. We also show the images of a Johannsen BH surrounded by a Page-Thorne thin accretion disk observed by a remote observer with a ray-tracing method and discuss the effects of the deviation parameters on deforming the accretion disk image, which could be tested by observations with higher sensitivities in the future.
\end{abstract}

\section{Introduction}
\label{section1}
Black hole may be the most thoroughly-studied object before its discovery in the history of science. Recently, the Event Horizon Telescope (EHT) collaboration announced the images of the supermassive BHs M87* and SgrA*\cite{EventHorizonTelescope:2019ggy,EventHorizonTelescope:2022xqj}, these discoveries together with the detections of gravitational waves (GWs)\cite{LIGOScientific:2016aoc,LIGOScientific:2016sjg} confirm the existence of black holes, and open a new era of testing gravity in strong-field regime and far beyond the scale of solar system.

Shadow is the image formed by photons which are sent off by the central BH and reach the observer finally.  All the unstable orbits of photons form light rings around BH, the projection of the light rings on the observing screen is what  called shadow. The unstable orbits are the critical ones, on which small perturbations would cause the photons to be either captured by the central compact object or sent off to infinity. The light rays forming shadow pass very close to the event horizon and are bent a lot due to the strong gravitational lensing effect, therefore observations of shadow make it possible to test the strong-field properties of gravity.

As we know, according to no-hair theorem\cite{Israel:1967wq,Carter:1971zc,Hawking:1971vc}, Kerr metric is the unique neutral, stationary, and asymptotically flat BH solution of vacuum Einstein equations in four dimensions. While, it is possible BHs are not isolated and are surrounded and influenced by other fields, as well as, the underlying gravity theory is possible to be some modified gravity theory, which means Kerr hypothesis may be be deviated.
Recently, non-Kerr BHs have been studied extensively by considering coupling matter fields to gravity\cite{Herdeiro:2014goa,Herdeiro:2016tmi,Bambi:2013ufa,Zubair:2022fiw}, or modifying theories of gravity \cite{Mureika:2016efo,Anson:2020trg,Chow:2010sf,Liu:2012xn,Ling:2021olm,BenAchour:2020fgy}, or immersing the BHs in astrophysical environment \cite{Babar:2021nst,Naji:2016srh,Das:2021otl,Jusufi:2020odz,Toshmatov:2015npp,Xu:2017bpz}.
Testing no-hair theorem, and thus testing general relativity, is one of the most important topics in black hole physics and astrophysics. 

 Observationally, it is quite sensible to treat the deviations of Kerr BH in a model-independent way, so that the effects of the deviations can be studies universally. The model-independent strong-field tests require a modified spacetime which deviates from the Kerr metric in a parametrized form. In Ref.\cite{Johannsen:2013szh}, Johannsen constructed a general asymptotically flat rotating BH that depends nonlinearly on the deviation functions, and the spacetime symmetries are held meanwhile. The Johannsen BH admits three constants of motion, i.e., the energy, angular momentum and Carter constant, which is regular outside the event horizon and  free of closed timelike curves. The BH possesses  correct Newtonian limit in the non-relativistic regime, and is consistent with all current weak-field tests.  Up to leading truncation order, the BH possesses four deviation parameters that measure potential deviations from the Kerr metric in the strong-field regime. Apparently, the BH is not solution of any particular theories of gravity, but can be embedded into several known modified theories of gravity invoked in astrophysics and cosmology by suitably choosing the deviation parameters.
Significantly Johannsen BH does not suffer from pathologies in the exterior domain,  it hence fairly serves as a
phenomenological framework for strong-field tests of the no-hair theorem in general classes of gravity theories.

The shadow of Schwarzschild BH was first studied by Synge\cite{Synge:1966okc}, and Bardeen first studied the shadow of Kerr\cite{Bardeen1973}, the result showed that, compared to the spherical BH, the shadow of Kerr is no longer circular, the distortion of shadow is understood as the effect of frame dragging. Aside from the distortion effects of rotating parameter, studies show that the deviation parameters lead to deformation of the shadows significantly as well, thus shadow can be used to extract information of BHs so as to detect the deviations from Kerr and constrain the BH parameters  in the coming observations\cite{Perlick:2015vta,Abdujabbarov:2015pqp,Konoplya:2019sns,Chang:2019vni,Li:2020drn,
EventHorizonTelescope:2020qrl,Atamurotov:2021hck,Chael:2021rjo,Badia:2021yrh,Briozzo:2022mgg,EventHorizonTelescope:2021dqv,Chakhchi:2022fls,Cunha:2016bjh,Kumar:2020hgm,Wei:2019pjf,Belhaj:2020rdb,Belhaj:2020okh,Gralla:2019xty,Abdikamalov:2019ztb,Okyay:2021nnh,Cimdiker:2021cpz,Kumaran:2022soh,Ovgun:2020gjz,Ovgun:2018tua,Herdeiro:2022yle,Cunha:2015yba,Cunha:2016bpi,Long:2020wqj,Afrin:2021imp}. The announcement of BH shadow inspire many attempts to test no-hair theorem, alternative theories of gravity, the candidates of dark matter, and quantum effects of gravity utilizing the data of shadow \cite{Abdujabbarov:2016hnw,Tsukamoto:2017fxq,Xavier:2023exm,Lambiase:2023hng,Badia:2020pnh,Paithankar:2023ofw,
Ling:2022vrv,Lima:2021las,Wei:2020ght,Wang:2021irh,Wang:2018prk,Guo:2020zmf,Kruglov:2020tes,Anacleto:2021qoe,Atamurotov:2021cgh,Wu:2021pgf,
Guo:2022nto,Kala:2022uog,Pantig:2022qak,Jha:2022nzd,Zeng:2022fdm,Gammon:2022bfu,Chang:2020miq,Liu:2020ola,Heydari-Fard:2021pjc,Heydari-Fard:2022xhr,Nozari:2023flq,Saghafi:2022pme,Volkel:2020xlc,Lara:2021zth,Shaikh:2021yux,Konoplya:2021slg,Ghosh:2023kge,
Tripathi:2018lhx,Tripathi:2020dni,Lee:2021sws,Rosa:2022tfv,Rosa:2023hfm,Olmo:2023lil,Rosa:2023qcv,Meng:2022kjs,Kuang:2022ojj,Tang:2022hsu,Wang:2023vcv,Meng:2023htc}.  With the improvements of detection accuracy, BH shadow is expected to serve as independent measurement or complementary to GW detections to provide a powerful way of constraining the BH parameters and distinguishing different BHs.

Aside from the deviation effects, the luminous source is another factor that affect the shadow. We know BHs accrete  matters and form accretion disk surrounding them, the matters in accretion disk would emit photons and serve as luminous source of the shadow, which is expected to be affected significantly by accretion disk. Compared to the ideal treatment of BH shadow, taking into account accretion disk is more realistic in astrophysics.

In this paper, we study the shadow of Johannsen BH, our results show that the shadow are affected significantly by rotating and deviation parameters, including the shape of shadow boundaries, and the luminosity, the central dark region and the region with high luminosity of the accretion disk images. With the observational results of M87* and SgrA* we constrain the deviation parameters further based on the theoretical constraints.
Due to the universal features of Johannsen BH, this study is helpful to test no-hair theorem and modified gravities in a model-independent way.

The paper is organised as, in section \ref{section2} we review the general rotating BH proposed by Johannsen. In section \ref{section3}, we plot the shadow boundaries on celestial plane, calculate the shadow radius, and constrain the deviation parameters with observational data. In section \ref{section4}, we give the accretion disk images of Johannsen BHs by ray-tracing method. We summarize our results in the last section \ref{section5}.
In this work,
the geometrized units are used with $G=c=1$.

\section{General rotating BH\label{section2}}
In this section we review Johannsen BH proposed in Ref.\cite{Johannsen:2013szh}. Johannsen BH is a general neutral, stationary and asymptoticall flat BH, which is a generalization of Kerr. Let's start with Kerr, in Boyer-Lindquist coordinates Kerr BH is given by
\begin{align}
ds^2&=-\left(1-\frac{2Mr}{\Sigma}\right)dt^2 -\frac{2Mar\sin^2\theta}{\Sigma}dtd\phi + \frac{\Sigma}{\Delta}dr^2\nonumber \\
&+ \Sigma d\theta^2+\left(r^2+a^2+\frac{2Ma^2r\sin^2\theta}{\Sigma}\right)\sin^2\theta d\phi^2,
\label{kerr}
\end{align}
with
\begin{align}
\Delta \equiv r^2-2Mr+a^2, \quad\quad\quad
\Sigma \equiv r^2+a^2\cos^2 \theta.
\label{deltasigma}
\end{align}
We know general stationary and axisymmetric BHs are of Petrov type I which admit two constants of motion, the energy $E$ and angular momentum $L_z$. While Kerr BH is of Petrov type D, since Carter found there exist a third constant of motion, the Carter constant $Q$\cite{Carter:1968rr},  by separating variables of the Hamilton-Jacobi equations of test particles
\begin{align}
  -\frac{\partial S}{\partial\tau} = \frac{1}{2} g^{\alpha\beta} \frac{\partial S}{\partial x^\alpha} \frac{\partial S}{\partial x^\beta}.\label{HJeq}
\end{align}
The successful separation of variables enable the geodesic motions of test particles in Kerr spacetime to be integrable. If the Kerr metric is required to be modified in the manner by holding the Carter symmetry, i.e., the  motions of test particle in this novel spacetime is expected to be integrable as well, a metric constructed by Johannsen satisfies this requirement, the metric can be expressed in contravariant form as
\begin{equation}
\begin{aligned}
& g^{\alpha\beta} \frac{\partial}{\partial x^\alpha} \frac{\partial}{\partial x^\beta}  \\
 =&  -\frac{1}{\Delta\tilde{\Sigma}} \left[ (r^2+a^2)A_1(r)\frac{\partial}{\partial t} + a A_2(r) \frac{\partial}{\partial \phi} \right]^2 \\
&+ \frac{\Delta}{\tilde{\Sigma}} A_5(r)\left( \frac{\partial}{\partial r} \right)^2 + \frac{1}{\tilde{\Sigma}} A_6(\theta)\left( \frac{\partial}{\partial \theta} \right)^2\\
&+ \frac{1}{\tilde{\Sigma}\sin^2\theta} \left[ A_3(\theta)\frac{\partial}{\partial \phi} + a \sin^2\theta A_4(\theta)\frac{\partial}{\partial t} \right]^2,
\label{controJohannsen}
\end{aligned}
\end{equation}
where
\begin{align}
\tilde{\Sigma} \equiv \Sigma + f(r) + g(\theta).
\end{align}
The explicit form of the deviation functions $A_i(r)$ and $A_j(\theta)$ will be discussed below.

\begin{figure*}
\begin{center}
\includegraphics[width=0.85 \textwidth]{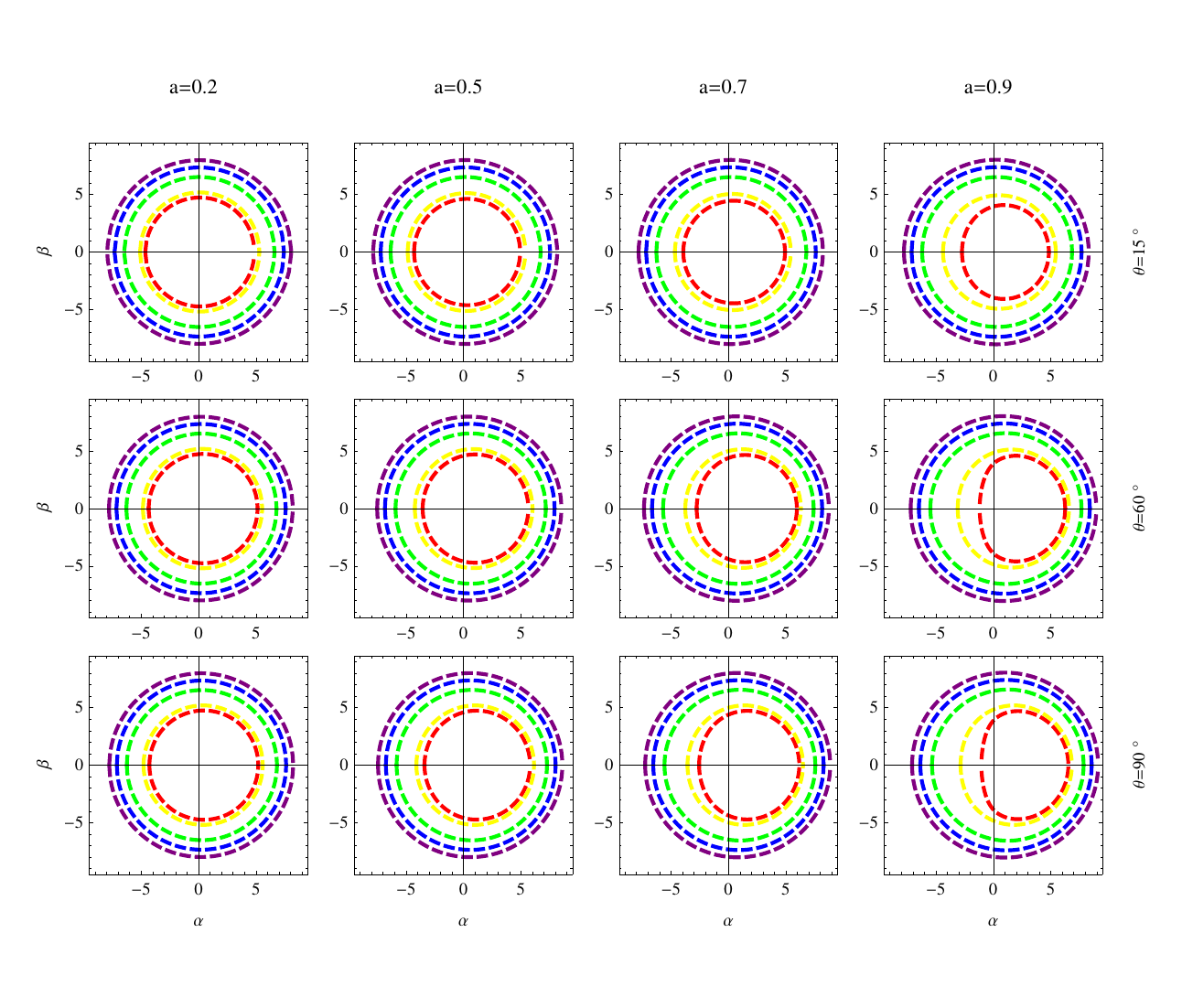}
\end{center}
\caption{Shadow boundaries of Johannsen BH with different values of $\alpha_{13}$, $a$ and $\theta_0$. All plots in the same row correspond to the same inclination angle. All plots in the same column correspond to the same rotating parameter. The red circle on every plot corresponds to $\alpha_{13}=-2$, the yellow circle corresponds to $\alpha_{13}=0$,  the green circle corresponds to $\alpha_{13}=10$, the blue circle corresponds to $\alpha_{13}=20$, and the purple circle corresponds to $\alpha_{13}=30$. Masses of the black holes are set to be unity $M=1$.}\label{ShadowBoundary}.
\end{figure*}

To see the motions of test particle in the spacetime described by the metric (\ref{controJohannsen}) are integrable,
we suppose the Hamilton-Jacobi function of a particle with mass $m_0$ to be of the form
\begin{align}
S=\frac{1}{2}m_0^2\tau-Et+L_z\phi+S_r(r)+S_\theta(\theta),
\end{align}
which indicates $p_r=\textmd{d} S_r/\textmd{d} r$ and $p_\theta=\textmd{d} S_\theta/\textmd{d} \theta$.
Inserting the metric (\ref{controJohannsen}) into the Hamilton-Jacobi equation (\ref{HJeq}), we have
\begin{equation}
\begin{aligned}
-m_0^2 = & -\frac{1}{\Delta\tilde{\Sigma}} \left[ -(r^2+a^2)A_1(r)E + a A_2(r)L_z \right]^2  \\
& + \frac{1}{\tilde{\Sigma}\sin^2\theta} \left[ L_z - a E \sin^2\theta \right]^2\\
&+ \frac{\Delta}{\tilde{\Sigma}} A_5(r)\left( \frac{\partial S_r}{\partial r} \right)^2 + \frac{1}{\tilde{\Sigma}}\left( \frac{\partial S_\theta}{\partial \theta} \right)^2.\label{HJeqsept}
\end{aligned}
\end{equation}
After separating variables of  Eq. (\ref{HJeqsept})  we have
\begin{align}
C = & \frac{1}{\Delta} \left[ -(r^2+a^2)A_1(r)E + a A_2(r)L_z \right]^2\nn\\
 &-m_0^2 \left[ r^2 + f(r) \right] - \Delta A_5(r)\left( \frac{\partial S_r}{\partial r} \right)^2,\label{HJeqsept1}\\
C = & \frac{1}{\sin^2\theta} \left[ L_z - aE \sin^2\theta \right]^2
 + m_0^2 (a^2 \cos^2 \theta+g(\theta))  + \left( \frac{\partial S_\theta}{\partial \theta} \right)^2,
\label{HJeqsept2}
\end{align}
where $C$ is the separation constant, through which we could define the Carter constant
\begin{align}
Q=C-(L_z- aE)^2.\label{QC}
\end{align}

Since Carter symmetry is a critical property of Johannsen metric, it is quite sensible to discuss the underlying physics of Carter symmetry/Carter constant. Carter constant reveals a deep and complicated symmetry in spacetime, it corresponds to a Killing tensor. For Johannsen BH, we find the Killing tensor is given by
      \be
      K^{\mu\nu}=a^2\cos^2\theta g^{\mu\nu}+\frac{1}{\sin^2\theta}\delta^\mu_\phi \delta^\nu_\phi+a^2\sin^2\theta \delta^\mu_t\delta^\nu_t-2a\delta^\mu_t\delta^\nu_\phi+\delta^\mu_\theta\delta^\nu_\theta.
      \en
       One checks that the Killing equation $K_{(\mu\nu;\lambda)}=0$ is satisfied, where the bracket denotes symmetrization of the indices. As Killing vectors $\partial_t$ and $\partial_\phi$ correspond to energy and angular momentum of the spacetime, Killing tensor describes a nontrivial symmetry of the spacetime and makes the general geodesic motion of probe particle to be integrable, for more motivations and applications of Carter constant see Cater's original paper and related works \cite{Carter:1977pq,Walker:1970un,Frolov:2017kze,deFelice:1999wm,Vigeland:2011ji}.

       Generally it relates to energy and angular momentum at $z$-direction. Here we concentrate on the physical implications of Cater constant in this article. The separation constant $C$ and Carter constant $Q$ in our paper are related through Eq.(\ref{QC}).
      We take new separation constant following \cite{deFelice:1999wm}
      \be
      \Lambda=C+2aEL_z-m_0^2a^2,
      \en
      then the separation equation (\ref{HJeqsept2}) can be rewritten as
      \be
      p_\theta^2+\frac{p_\phi^2}{\sin^2\theta}+m_0^2a^2\left[\left(\frac{E}{m_0}\right)^2-1\right]\sin^2\theta=\Lambda,\label{sepeq}
      \en
      where $p_\phi=L_z$. Since we will see later $g(\theta)$ is set to be zero by observation requirement, here we ignore this term to clarify the physical meaning of Carter constant. Note that eq.(\ref{sepeq}) does not depend on the mass $M$ of gravitational source. Now we try to explain the terms on the left side of above equation. In the $a\rightarrow0$ limit, i.e., the spherical symmetry case, the physical meaning of the first two terms are clear. They are the square of angular momentum $L^2=p_\theta^2+\frac{p_\phi^2}{\sin^2\theta}$, in which the separation constant $\Lambda$ equals to $L^2$.
      When $a\neq0$, we consider the weak-field approximation, which implies that the observer locates at far field region, the spatial metric can be rewritten as
      \be
      d\sigma^2=\frac{\Sigma}{r^2+a^2}dr^2+\Sigma d\theta^2+(r^2+a^2)\sin^2\theta d\phi^2,\label{spametric}
      \en
      with oblate spheroidal coordinates
      \bea
           x&=&(r^2+a^2)^{1/2}\sin\theta\cos\phi,\\
      y&=&(r^2+a^2)^{1/2}\sin\theta\sin\phi,\\
      z&=&r\cos\theta.
      \ena
      With the asymptotic metric (\ref{spametric}) the square modulus of particle's linear momentum observed at infinity is given by
      \be
      p^2=g^{rr}_{\infty}p_r^2+g^{\theta\theta}_{\infty}p_\theta^2+g^{\phi\phi}_{\infty}p_\phi^2.
      \en
      In the limit $r\rightarrow\infty$, we have
      \be
      g^{\theta\theta}_{\infty}p_\theta^2\longrightarrow0,\quad\quad\quad\quad
      g^{\phi\phi}_{\infty}p_\phi^2\longrightarrow0.
      \en
      Thus
      \be
      g^{rr}_{\infty}p_r^2\longrightarrow p^2=m_0^2\left[\left(\frac{E}{m_0}\right)^2-1\right].
      \en
      Now the separation equation (\ref{sepeq}) can be rewritten as
      \be
      p_\theta^2+\frac{p_\phi^2}{\sin^2\theta}+|p_r|^2a^2\sin^2\theta=\Lambda.\label{sepeq2}
      \en
     This equation implies that, if a particle at infinity possess nonvanishing momentum $p_r$ it also possesses nonvanishing angular momentum with respect to the center $O$ of $r=0$ disk. In the Boyer-Lindquist coordinates, the $\theta=constant$ surfaces are not cones with vertices at the origin, but are hyperboloids of rotation, which cross the $r=0$ disk into a circle with radius $a\sin\theta$. At infinity the vector $\vec{p}_r$ coincide with the asymptotes of the hyperboloid, which is to say $\vec{p}_r$ is tangent to $\theta=constant$ surfaces, thus parallel transport along $\theta=constant$ surface maintains $\vec{p}_r$ to be tangent to the hyperboloid. When crossing the $r=0$ disk, the point of maximum approach to the center $O$  is reached, now $\vec{p}_r$ is orthogonal to the $r=0$ disk. So its momentum with respect to the center $O$ has a square modulus given by $|p_r|^2a^2\sin^2\theta$, which is just the term appearing in eq.(\ref{sepeq2}). It should be mentioned, for bounded orbits of particles the sign before the third term on the left side of eq.(\ref{sepeq2}) should be changed, for more details please refer to Ref.\cite{deFelice:1999wm}.

     From eq.(\ref{sepeq2}) we see that, the separation constant has the meaning of `extended' angular momentum which come not only from the angular motions but also from the radial motion. The radial angular momentum is not obvious, since for the spacetime of black hole with spherical symmetry, angular momentum comes only from angular motions. While, for the spacetime of rotating black hole, the radial angular momentum appears. It arises from the particular choice of coordinates which in turn ensures the separability of the Hamilton-Jacobi equation.

Solving the variables-separated Eqs. (\ref{HJeqsept1}) and (\ref{HJeqsept2}) we have
\begin{align}
S_r(r) = \pm \int dr \frac{1}{\Delta} \sqrt{ \frac{ R(r) }{ A_5(r) } },\;
S_\theta(\theta) = \pm \int d\theta \sqrt{ \Theta(\theta) },
\end{align}
with
\begin{align}
R(r) &\equiv P^2
- \Delta \left\{ m_0^2 \left[ r^2 + f(r) \right] + (L_z - aE)^2 + Q \right\},\label{Rr}
\\
\Theta(\theta) &\equiv Q + (L_z-aE)^2 -m_0^2 (a^2 \cos^2 \theta+g(\theta))
- \frac{1}{\sin^2\theta} \left[ L_z - a E \sin^2\theta \right]^2, \\
P &\equiv (r^2 + a^2)A_1(r)E - aA_2(r) L_z.\label{Pr}
\end{align}
The explicit integration expression of coordinates and proper time can be obtained by setting the partial derivative of Hamilton-Jacobi function with respect to the constants of motion to be zero, that's to say the motions of particles in spacetime (\ref{controJohannsen}) are integrable.

\begin{figure*}[h]
\begin{center}
\includegraphics[width=0.85 \textwidth]{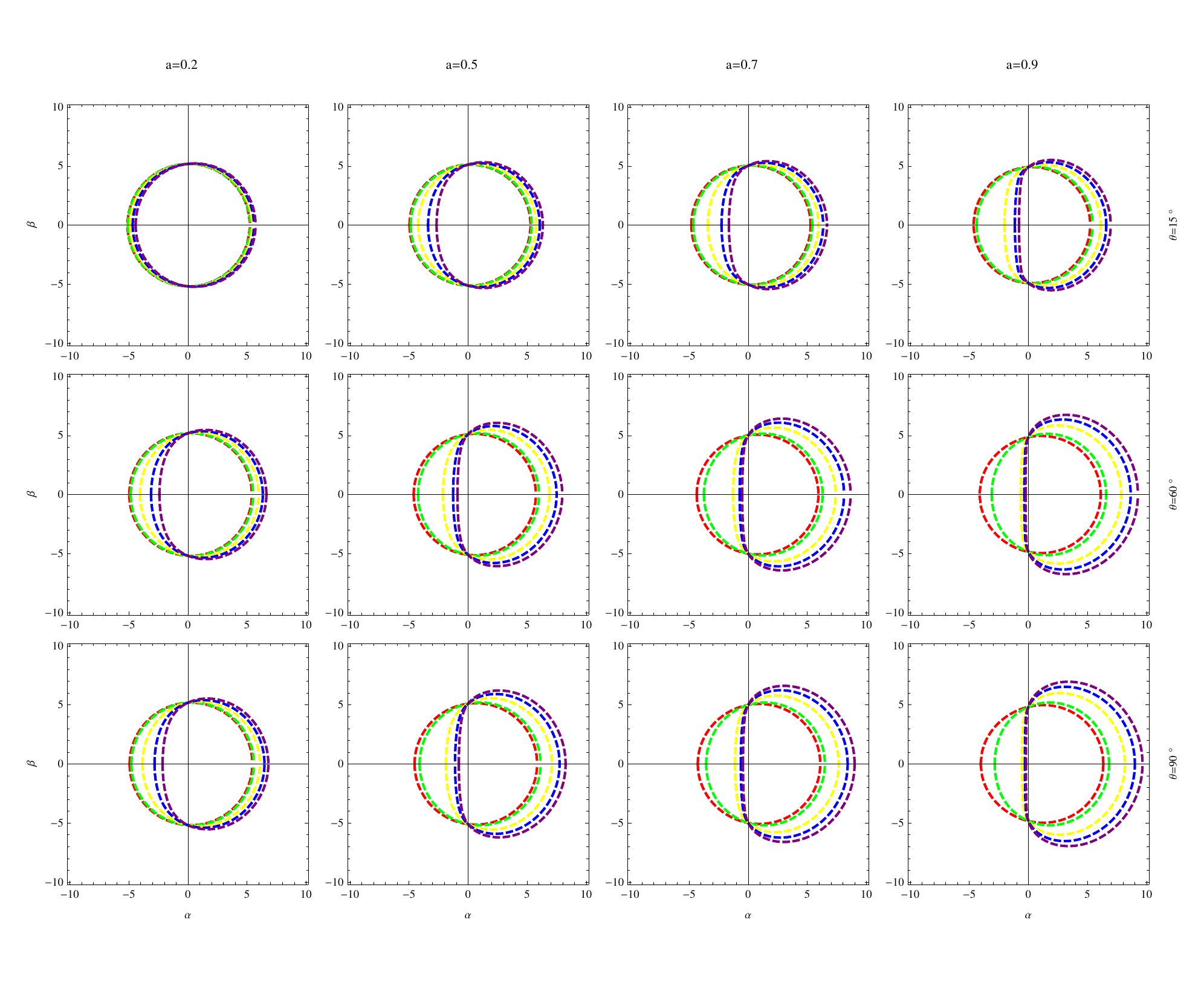}
\end{center}
\caption{Shadow boundaries of Johannsen BH with different values of $\alpha_{22}$, $a$ and $\theta_0$. All plots in the same row correspond to the same inclination angle. All plots in the same column correspond to the same rotating parameter. The red circle on every plot corresponds to $\alpha_{22}=-2$, the green circle corresponds to $\alpha_{22}=0$,  the yellow circle corresponds to $\alpha_{22}=10$, the blue circle corresponds to $\alpha_{22}=20$, the purple circle corresponds to $\alpha_{22}=30$. Masses of the black holes are set to be unity $M=1$.}\label{ShadowBoundary2}
\end{figure*}

In order to write the metric in explicit form, the derivation functions $A_i(r)$, $i=1,2,5$, can be expanded as a power series of $M/r$\cite{Johannsen:2013szh}:
\begin{align}
A_i(r) \equiv \sum_{n=0}^\infty \alpha_{in} \left( \frac{M}{r} \right)^n,~~~~i=1,2,5,
\end{align}
as well as
\begin{align}
f(r) \equiv \sum_{n=0}\epsilon_n \frac{M^n}{r^{n-2}}, \;\;
g(\theta) \equiv M^2 \sum_{k,l=0}^\infty \gamma_{kl} \sin^k\theta \cos^l\theta,
\end{align}
Expanding the metric in power of $1/r$ and requiring the metric to be asymptotically flat, the undetermined parameters and functions are fixed to be $\alpha_{10}=\alpha_{20}=\alpha_{50}=1$, $\epsilon_0=\epsilon_1=0$, and $A_3(\theta)=A_4(\theta)=A_6(\theta)=1$. If the parameter $M$ is required to be the mass of the central object, the parameters can be further constrained to be $\alpha_{11}=\alpha_{21}=\alpha_{51}=0$. The metric should be consistent with the the weak-field tests, in the parameterized post-Newtonian (PPN) formulism, the general metric can be written in the form \cite{Will:2005va,WillBook}
      \be
      ds^2=-\left[1-\frac{2M}{r}+2(\beta_{PPN}-\gamma_{PPN})\frac{M^2}{r^2}\right]dt^2+(1+2\gamma_{PPN}\frac{M}{r})dr^2+r^2(d\theta^2+\sin^2\theta d\phi^2).\label{metricPPN}
      \en
 For general relativity, $\beta_{PPN}=\gamma_{PPN}=1$. Values of $\beta_{PPN},\gamma_{PPN}$ corresponding to some other metric theories can be found in Chapter 5 of Ref.\cite{WillBook}.
We perform the large $r$ expansion of Johannsen metric  and find
\be
      ds^2=-\left[1-\frac{2M}{r}-\frac{M^2(2\alpha_{12}-\epsilon_2)-g(\theta)}{r^2}\right]dt^2
      +(1+\frac{2M}{r})dr^2+r^2(d\theta^2+\sin^2\theta d\phi^2).\label{metricexpand}
      \en
Note that the spin parameter $a$ appear in the higher order expansion of $\frac{1}{r}$. Comparing metric (\ref{metricexpand}) with metric (\ref{metricPPN}) we have
\bea
2(\beta_{PPN}-\gamma_{PPN})&=&\epsilon_2-2\alpha_{12}+\frac{g(\theta)}{M^2},\\
\gamma_{PPN}&=&1,
\ena
which implies
\be
\beta_{PPN}-1=\frac{1}{2}\left[\epsilon_2-2\alpha_{12}+\frac{g(\theta)}{M^2}\right]
\en
This quantity is constrained by observations to be \cite{WillBook}
\be
\beta_{PPN}-1=(0.2\pm 2.5) \times 10^{-5}.
\en
In order to avoid any fine-tuning between the parameters
$\epsilon_2$, $\alpha_{12}$ and the function $g(\theta)$, one can set $\epsilon_2=\alpha_{12}=g(\theta)=0$ for simplicity.

  Now the metric can be expressed in covariant form as
\begin{equation}
\begin{aligned}
g_{tt} &= -\frac{\tilde{\Sigma}[\Delta-a^2A_2(r)^2\sin^2\theta]}{[(r^2+a^2)A_1(r)-a^2A_2(r)\sin^2\theta]^2},  \\
g_{t\phi} &= -\frac{a[(r^2+a^2)A_1(r)A_2(r)-\Delta]\tilde{\Sigma}\sin^2\theta}{[(r^2+a^2)A_1(r)-a^2A_2(r)\sin^2\theta]^2}, \\
g_{rr} &= \frac{\tilde{\Sigma}}{\Delta A_5(r)}, \label{covmetric} \\
g_{\theta \theta} &= \tilde{\Sigma}, \\
g_{\phi \phi} &= \frac{\tilde{\Sigma} \sin^2 \theta \left[(r^2 + a^2)^2 A_1(r)^2 - a^2 \Delta \sin^2 \theta \right]}{[(r^2+a^2)A_1(r)-a^2A_2(r)\sin^2\theta]^2}.\nn
\end{aligned}
\end{equation}
with
\begin{equation}
\begin{aligned}
A_1(r) &= 1 + \sum_{n=3}^\infty \alpha_{1n} \left( \frac{M}{r} \right)^n,\\
A_2(r) &= 1 + \sum_{n=2}^\infty \alpha_{2n} \left( \frac{M}{r} \right)^n,\\
A_5(r) &= 1 + \sum_{n=2}^\infty \alpha_{5n} \left( \frac{M}{r} \right)^n,\\
\tilde{\Sigma} &= r^2 + a^2 \cos^2\theta + f(r),\\
f(r) &= \sum_{n=3}^\infty\epsilon_n \frac{M^n}{r^{n-2}}.
\label{devfuncs}
\end{aligned}
\end{equation}
Thus to leading order, the metric is characterized by mass $M$, spin $a$ and four deviation parameters $\alpha_{13}$, $\alpha_{22}$, $\alpha_{52}$, and $\epsilon_3$. By requiring the metric to be regular outside the event horizon, the determinant of the metric to be non-negative, and the absence  of closed timelike curves, the four deviation parameters can be further constrained to be
\begin{equation}
\begin{aligned}
\alpha_{52} &> - \frac{ \left( M+\sqrt{M^2-a^2} \right)^2 }{M^2},\\
\epsilon_3 &> - \frac{ \left( M+\sqrt{M^2-a^2} \right)^3 }{M^3},\\
\alpha_{13} &> - \frac{ \left( M+\sqrt{M^2-a^2} \right)^3 }{ M^3 },\\
\alpha_{22} &> - \frac{ \left( M+\sqrt{M^2-a^2} \right)^2 }{ M^2 }.\label{theoryconstraint}
\end{aligned}
\end{equation}
For more details on constraining the deviation parameters from theoretical aspect please refer to Ref.\cite{Johannsen:2013szh}. It should be mentioned that, generally the Johannsen metric (\ref{covmetric}) is not a solution of any particular gravity theory, but can be mapped to known four-dimensional BH solution of theories of modified gravity by suitably choosing the deviation parameters, e.g., the metric can be mapped to modified gravity bumpy Kerr metric, rotating braneworld BHs, slowly rotating BHs in dynamical Chern-Simons gravity, and static BHs in Einstein-Dilaton-Gauss-Bonnet gravity\cite{Vigeland:2011ji,Yunes:2011we,Yunes:2009hc,Aliev:2005bi}.

\begin{figure*}
\begin{center}
\includegraphics[width=0.45 \textwidth]{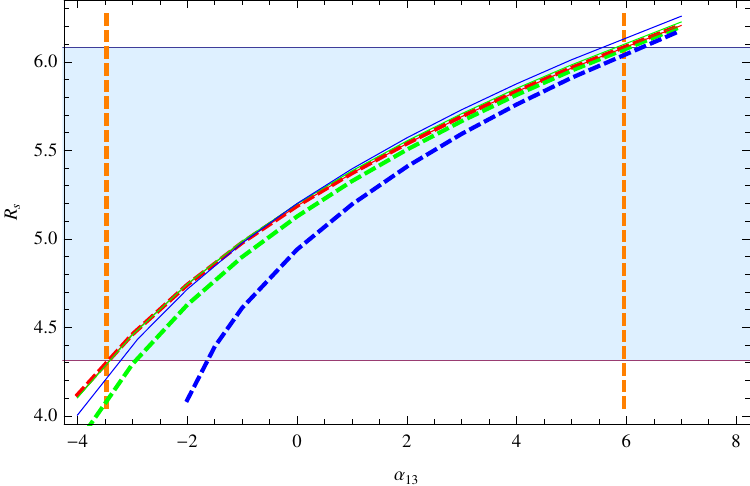}
\includegraphics[width=0.45 \textwidth]{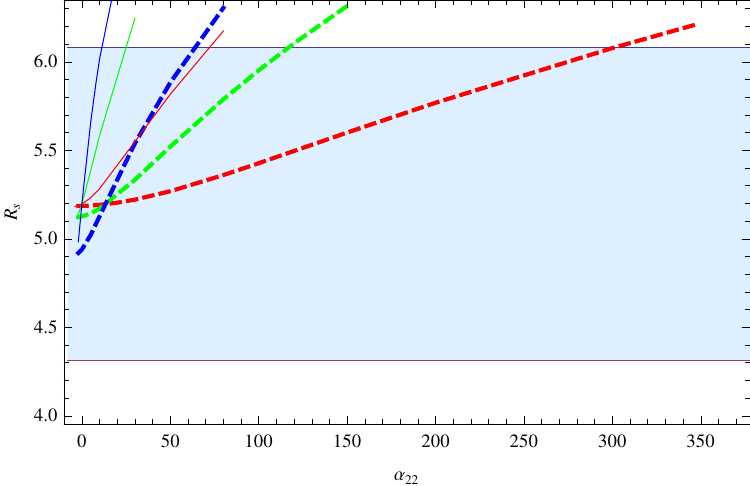}
\end{center}
\caption{Shadow radius as function of $\alpha_{13}$ (left plot) or $\alpha_{22}$ (right plot). The dashed and solid lines correspond to $\theta_0=15^{\circ}$ and $90^{\circ}$ respectively. The red, green and blue lines correspond to $a=0.2, 0.5$ and 0.9 respectively. The  blue region on each plot represents the observation bound of M87*. The orange vertical lines represent  the  case $a=0.2$ and $\theta_0=15^{\circ}$. Masses of the black holes are set to be unity $M=1$.}\label{constraint}
\end{figure*}

\section{Shadow boundary\label{section3}}
In this section, we study the effects of the BI parameters on the shadow boundary. Shadow boundary is expected to be formed by all the unstable orbits of photons known as light rings. The unstable orbits are the critical ones which separate the photons to escape to infinity or to be captured by the central BH.
\subsection{Shadow boundary on celestial plane}
With the results given in the last section, one obtains the geodesic equations of photon in the spacetime of Johannsen BH
\begin{equation}\label{eq_geo}
\begin{aligned}
\tilde{\Sigma} \frac{dt}{d\tau} &= -a \left[aE\sin^2 \theta -L_z\right]
 + \frac{(r^2+a^2)A_1(r)}{\Delta} P,\\
\tilde{\Sigma} \frac{dr}{d\tau} &= \pm \sqrt{A_5(r) R(r)},\\
\tilde{\Sigma} \frac{d\theta}{d\tau} &= \pm \sqrt{\Theta(\theta)},\\
\tilde{\Sigma} \frac{d\phi}{d\tau} &= -\left[ aE -\frac{L_z}{\sin^2 \theta} \right]
+ \frac{aA_2(r)}{\Delta} P.
\end{aligned}
\end{equation}
The unstable circular  orbit is determined by the equations
\begin{align}
R=0,\quad\quad\quad\quad\quad\quad\frac{\textmd{d}R}{\textmd{d}r}=0.\label{Eqboundary}
\end{align}
Solving the above equation we have
\bea
L_z&=&E\left[3a^4M^3\alpha_{13}-a^2r(Mr^3+r^4+7M^4\alpha_{13}-5M^3r\alpha_{13})-r^3(-3Mr^3+r^4+3M^4
\alpha_{13}\right. \nonumber\\
&~&\quad\left.-2M^3r\alpha_{13})\right]/\left[ar(-M r^3+r^4-5M^3r\alpha_{22}+M^2(2a^2+3r^2)\alpha_{22})\right],\label{solLz}\\
Q&=&(E^2(-r^{10}(-3Mr^3+r^4+3M^4\alpha_{13}-2M^3r\alpha_{13})^2+a^8M^7\alpha_{13}\alpha_{22}
(6Mr^2\alpha_{13}\nonumber\\
&\quad &-4r^3\alpha_{22}+M^3\alpha_{13}\alpha_{22})-a^6M^4r(12M^5r^2\alpha_{13}^2\alpha_{22}+6Mr^6(\alpha_{13}-
2\alpha_{22})\alpha_{22}\nonumber\\
&\quad &-8r^7\alpha_{22}^2+2M^7\alpha_{13}^2\alpha_{22}^2+M^6r\alpha_{13}^2\alpha_{22}^2-2M^4r^3\alpha_{13}\alpha_{22}
(\alpha_{13}+4\alpha_{22})+M^2r^5\alpha_{13}\nonumber\\
&\quad &(15\alpha_{13}+34\alpha_{22})+8M^3r^4\alpha_{13}(-3\alpha_{13}-\alpha_{22}^2))+a^2Mr^5(4r^{10}-2M^2r^8(7
\alpha_{13}-2\alpha_{22})\nonumber\\
&\quad &+10Mr^9\alpha_{22}-2M^{10}\alpha_{13}^2\alpha_{22}^2+M^9r\alpha_{13}^2\alpha_{22}^2-M^5r^5\alpha_{13}(19\alpha_{13}+
30\alpha_{22})-2M^3r^7\nonumber\\
&\quad &(24\alpha_{13}+(15-8\alpha_{22})\alpha_{22})-2M^7r^3\alpha_{13}(8\alpha_{22}^2+\alpha_{13}(21+\alpha_{22}))+2M^6r^4\alpha_{13}
(28\alpha_{13}\nonumber\\
&\quad &+\alpha_{22}(15+4\alpha_{22}))+2M^4r^6(-16\alpha_{22}^2+\alpha_{13}(21+4\alpha_{22})))+a^4M^2r^3(8r^9\alpha_{22}\nonumber\\
&\quad &+8M^7r^2\alpha_{13}^2\alpha_{22}+2M^3r^6(\alpha_{13}-\alpha_{22})\alpha_{22}+4M^9\alpha_{13}^2\alpha_{22}^2-M^8r
\alpha_{13}^2\alpha_{22}^2+2Mr^8\nonumber\\
&\quad &(5\alpha_{13}+8\alpha_{22})-5M^4r^5(6\alpha_{13}^2+8\alpha_{13}\alpha_{22}+5\alpha_{22}^2)+M^2r^7(-22\alpha_{13}+23\alpha_{22}^2)\nonumber\\
&\quad &+2M^5r^4\alpha_{13}(38\alpha_{13}+\alpha_{22}(35+2\alpha_{22}))+M^6r^3\alpha_{13}(8\alpha_{22}^2-\alpha_{13}(49+6\alpha_{22})))))/\nonumber\\
&\quad &(a^2r^6(-Mr^3+r^4-5M^3r\alpha_{22}+M^2(2a^2+3r^2)\alpha_{22})^2).\label{solQ}
\ena

We introduce celestial coordinates to visualize the shadow of BH\cite{Abdujabbarov:2016hnw,Hioki:2009na}
\begin{equation}
 \begin{aligned}
\alpha&=\lim_{r_0\rightarrow\infty}\left(-r_0^2\sin\theta_0\frac{d\phi}{dr}\right)=-\xi\csc\theta_0,\\
\beta&=\lim_{r_0\rightarrow\infty}\left( r_0^2\frac{d\theta}{dr}\right)=\pm\left(\eta+a^2\cos^2\theta_0-\xi^2\cot^2\theta_0\right)^{1/2},
\end{aligned}\label{celestialcoord}
\end{equation}
where $r_0$ is the distance between the BH and the observer,  $\theta_0$ is the inclination angle between the rotating axis of the BH and the observer's line of sight. $\xi$ and $\eta$ are two new parameters which are introduced as $\xi=L_z/E$ and $\eta=Q/E^2$. For the $\theta=\pi/2$ plane, the celestial coordinates take the simple form: $\alpha=-\xi, \beta=\pm \sqrt{\eta}$.

From  Eqs.(\ref{solLz}) and (\ref{solQ}) we see that the shadow boundaries are not affected by the deviation parameters $\epsilon_3$ and $\alpha_{52}$, but are only affected by $\alpha_{13}$ and $\alpha_{22}$.  In Fig.\ref{ShadowBoundary} and Fig.\ref{ShadowBoundary2}, we plot the shadow boundaries of Johannsen BH with different values of $\alpha_{13}$ and $\alpha_{22}$.  It's interesting to see the effects of BH parameters by comparing Fig.\ref{ShadowBoundary} and Fig.\ref{ShadowBoundary2}. From the two figures we see that, the shadow size grow with both  $\alpha_{13}$ and $\alpha_{22}$. Fig.\ref{ShadowBoundary2} exhibits the shadow size grow explicitly with $a$ and $\theta_0$ as well, while  the effects of shadow size growing with $a$ and $\theta_0$ are not so evident seen from Fig.\ref{ShadowBoundary}. This is in agreement with the direct calculations of shadow radii  as shown in Fig.\ref{constraint} and Fig.\ref{constraint2},  in Fig.\ref{constraint} the curves are closer together while in Fig.\ref{constraint2} the curves are more widely separated. Actually, as we will discuss below, there exist intersection points for the curves in Fig.\ref{constraint} and Fig.\ref{constraint2}, on the two sides of intersection points the behaviors of shadow size are distinct. If the shadow size grow with $a$ and $\theta_0$ on one side of the intersection point, it shrink with $a$ and $\theta_0$ on the other side of the intersection point, and vice versa.  Fig.\ref{ShadowBoundary} and Fig.\ref{ShadowBoundary2} show that shadows are  deformed more greatly by larger $a$ and $\theta_0$, and $\alpha_{13}$, $\alpha_{22}$ contribute oppositely on deforming the shadows, the shadows are deformed more greatly by smaller $\alpha_{13}$ but are more deformed by larger $\alpha_{22}$.

\subsection{Constrain the BH parameters with observations}
Aside from the theoretical constraints on BH parameters given by Eq.(\ref{theoryconstraint}), the two relevant parameters that affect shadow boundaries, $\alpha_{13}$ and  $\alpha_{22}$, can be further constrained by observation data. For instance, the shadow radius observed by EHT collaboration can be used to constrain the BH parameters. For rotating BHs, the shadows of which are often not round, thus the definition of shadow radius for static BHs is not applicable here, we adopt the definition of shadow radius given in Refs.\cite{Hioki:2009na,Ling:2022vrv}
\begin{align}
R_s=\frac{\left(\alpha_t-\alpha_r\right)^2+\beta_t^2}{2(\alpha_r-\alpha_t)}.
\end{align}
Here $(\alpha_t, \beta_t)$, $(\alpha_r, 0)$ are coordinates of the shadow vertices at top and right edges,
which depend on the BH parameters through Eqs.(\ref{Eqboundary}) and (\ref{celestialcoord}), therefore the shadow radii would be affected by the BH parameters.

In this paper, we adopt the two observation results on shadow size of M87* and SgrA* released by EHT collaboration. For M87*, the observed angular gravitational radius reported by EHT collaboration is consistent within the range of $17\%$ with the one that was estimated through the analysis of stellar dynamics\cite{EventHorizonTelescope:2019ggy}, which means that the shadow radius is bounded as \cite{Xavier:2023exm}
\begin{align}
  4.31 M \leq R_s \leq 6.08 M. \label{boundM87}
\end{align}
The estimates on mass-to-distance ratio of SgrA* were predominantly produced by two teams of EHT collaboration, Keck Observatory and Very Large Telescope Interferometer (VLTI)\cite{EventHorizonTelescope:2022xqj}. The shadow size bound given by Keck is
\begin{align}
  4.5 M \leq R_s \leq 5.5 M. \label{boundKeck}
\end{align}
and the one given by VLTI is
\begin{align}
  4.3 M \leq R_s \leq 5.3 M. \label{boundKeck}
\end{align}

\begin{figure*}
\begin{center}
\includegraphics[width=0.45 \textwidth]{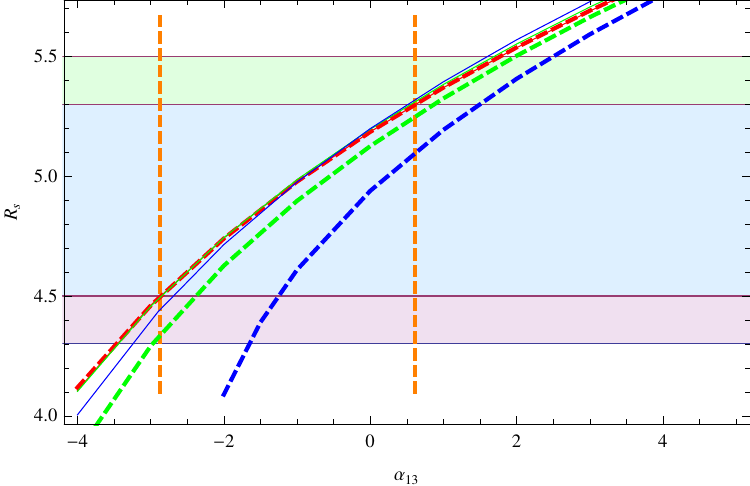}
\includegraphics[width=0.45 \textwidth]{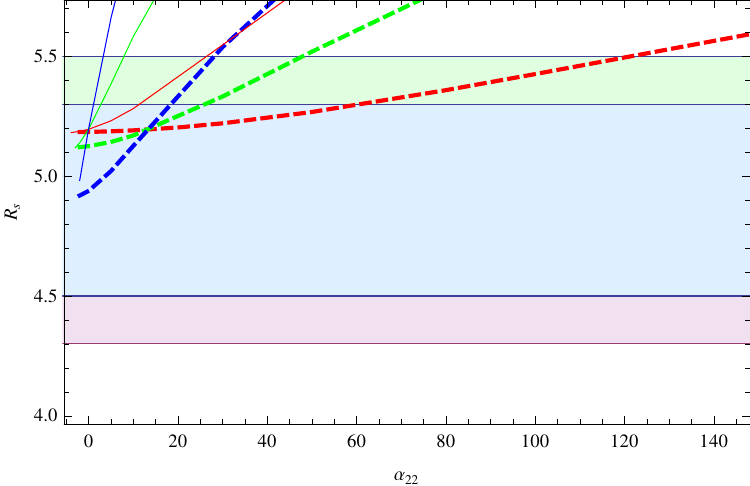}
\end{center}
\caption{Shadow radius as function of $\alpha_{13}$ (left plot) or $\alpha_{22}$ (right plot). The dashed and solid lines correspond to $\theta_0=15^{\circ}$ and $90^{\circ}$ respectively. The red, green and blue lines correspond to $a=0.2, 0.5$ and 0.9 respectively. The green region represents the observation bound of SgrA* given by VLTI, the purple region represents the observation bound of SgrA* given by Keck, the blue region is the intersection sector of the green and purple regions.  The orange vertical lines represent  the  case $a=0.2$ and $\theta_0=15^{\circ}$. Masses of the black holes are set to be unity $M=1$. }\label{constraint2}
\end{figure*}

In Fig.\ref{constraint} we plot shadow radii with different values of $a$,  $\theta_0$ and deviation parameters. For the left plot we fix $\alpha_{22}=0$ with $\alpha_{13}$ nonvanishing, and for the right plot we fix $\alpha_{13}=0$ with $\alpha_{22}$ nonvanishing. The blue regions describe the observation bound of M87*. Fig.\ref{constraint} exibits explicitly that, the shadow radii grow with the deviation parameters, within the theoretical allowed ranges of BH parameters, the theoretical results may exceed the observation bound. On the left plot, one sees the theoretical results of shadow radius may exceed both the upper and the lower observation bounds, i.e., too large and too small values of $\alpha_{13}$ are all disfavored. On the right plot, one sees that the theoretical shadow radius corresponding to the lower theoretical bound of $\alpha_{22}$ always lie in the range of observation bound, as $\alpha_{22}$ increases the upper observation bound may be exceeded, which implies too large values of $\alpha_{22}$ are disfavored. To see the observation constraints on the deviation parameters more explicitely, we plot two orange vertical lines for the  case $a=0.2$ and $\theta_0=15^{\circ}$ which is described by red dashed line. The values of $\alpha_{13}$ between the two orange vertical lines are preferred by observation, while the values of $\alpha_{13}$ beyond this range are disfavored by observation.

From Fig.\ref{constraint} one also sees that, for larger rotating parameter the shadow radii grow faster with the deviation parameters, this is in agreement with what one sees in Figs.\ref{ShadowBoundary} and \ref{ShadowBoundary2}. Since the growth rates of the curves are different for different $a$, there exist intersection points between the curves as shown in Fig.\ref{constraint}. On the left side of the intersection points, the shadow radii are larger for smaller rotating parameters with fixed deviation parameter, while on the right side of the intersection points, the shadow radii are smaller for smaller rotating parameters.
The effects of the rotating parameter are less significant for larger inclination angle, e.g., when $\theta_0=90^{\circ}$ the three curves with different $a$ almost coincide as shown on the left plot in Fig.\ref{constraint}.

In Fig.\ref{constraint2}, we use the data of SgrA* shadow to constrain the BH parameters. The green region corresponds to the observation bound given by Keck, the purple region corresponds to the observation bound given by VLTI, and the blue region is the overlapping region of the green and purple regions. From the figure it's easy to see,  too large and too small values of  $\alpha_{13}$ are disfavored likewise, so as to the theoretical shadow radius does not exceed the observation bounds. Too large values of $\alpha_{22}$ are disfavored since the observation upper bounds may be exceeded. While for the theoretical allowed minimum of $\alpha_{22}$, the corresponding shadow radii always lie in the range of the observation bounds.
Likewise, we also plot two orange vertical lines to indicate the observation preferred range of the deviation parameter for the case $a=0.2$ and $\theta_0=15^{\circ}$ corresponding to the red dashed line.  Now, the values of $\alpha_{13}$ between the two orange vertical lines are preferred by both  Keck  and VLTI bounds.

\section{The image of an accretion disk of Johannsen BH\label{section4}}
BHs accrete surrounding matter to form the accretion disk, which is expected to be an important luminous source of shadow, thus the information of accretion disk can be extracted through observing photons on the image plane of observers. We simulate the photons emitted from accretion disk by tracing backward along the trajectory of photon from the image plane to the accretion disk plane. The radiation flux from accretion disk corresponds to the luminosity on the image plane. Since accretion disks are quite complex physical systems which are difficult to simulate, for the purpose of this work, we use the simple Page-Thorne thin disk model to describe the radiation flux of accretion disk\cite{Page:1974he}.

\begin{figure}[h]
\begin{center}
\includegraphics[width=0.6 \textwidth]{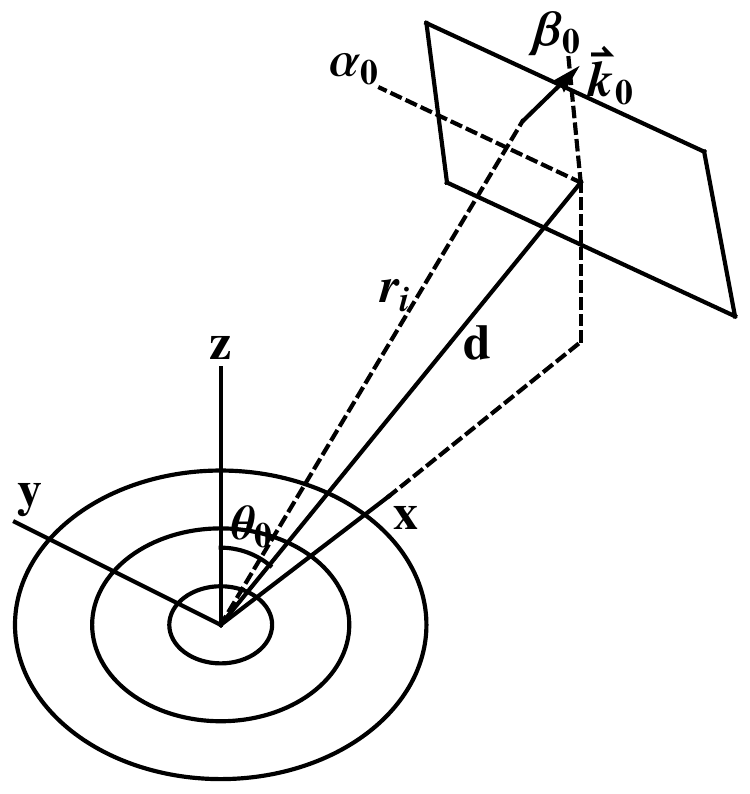}
\end{center}
\caption{The geometry used for tracing  trajectories of photons.}\label{obsfig}.
\end{figure}

The Lagrangian of a free particle is given by
\begin{equation}
{\cal L} = \frac{1}{2} g_{\alpha \beta} \dot{q}^\alpha \dot{q}^\beta,
\end{equation}
where $\dot{q}^\alpha\equiv \frac{dq^\alpha}{d\tau}$. The Hamiltonian could be defined in the standard way
\begin{equation}
H = p_\sigma \dot{q}^\sigma - {\cal L}
= \frac{1}{2}g^{\alpha \beta} p_\alpha p_\beta,
\end{equation}
where $p_\alpha\equiv \frac{\partial {\cal L}}{\partial\dot{q}^\alpha}= g_{\alpha \beta} \dot{q}^\beta$ is the momentum conjugate to $q^\alpha$. For the Johannsen metric we obtain
\begin{align}
H=\frac{\Delta A_5(r)}{2\tilde{\Sigma}}p_r^2+\frac{p_\theta^2}{2\tilde{\Sigma}}-\frac{R(r)}{2\Delta\tilde{\Sigma}}-\frac{\Theta}{2\tilde{\Sigma}},
\end{align}
from which we obtain the equations of motion of photon
\begin{equation}
\begin{aligned}
\dot{r} & = \frac{\Delta A_5(r)}{\tilde{\Sigma}}p_{r},\\
\dot{p}_{r} & =
-\left(\frac{\Delta A_5(r)}{2\tilde{\Sigma}}\right)'p_{r}^{2} -
\left(\frac{1}{2\tilde{\Sigma}}\right)'p_{\theta}^{2} + \left(\frac{R +
\Delta\Theta}{2\Delta\tilde{\Sigma}}\right)',\\
\dot{\theta}& =  \frac{p_{\theta}}{\tilde{\Sigma}},\\
\dot{p}_{\theta} & =
-\left(\frac{\Delta A_5(r)}{2\tilde{\Sigma}}\right)_{,\theta}p_{r}^{2} -
\left(\frac{1}{2\tilde{\Sigma}}\right)_{,\theta}p_{\theta}^{2} + \left(\frac{R +
\Delta\Theta}{2\Delta\tilde{\Sigma}}\right)_{,\theta},\\
\dot{t} & =  \frac{1}{2\Delta\tilde{\Sigma}} \frac{\partial}{\partial E} \left(R +\Delta\Theta \right),\\
\dot{p_t} & =  0,  \\
\dot{\phi} & =  -\frac{1}{2\Delta\widetilde{\Sigma}}\frac{\partial}{\partial L} \left(R + \Delta\Theta \right),\\
\dot{p_\phi} & =  0,
\end{aligned}\label{eomphon}
\end{equation}
where $'$ denotes derivation with respect to $r$, and the subscript $,\theta$ denotes derivation with respect to $\theta$. We consider an observer viewing the central BH from a large distance $d$ (see Fig.\ref{obsfig}). According to the equations of motion (\ref{eomphon}), the photons with initial positions and momenta on the observer's image plane can be traced backwards to the accretion disk so as to extract information of the accretion disk. For convenience, we convert the coordinates of the photon  on the image plane at $(\alpha_0, \beta_0)$  to Boyer-Lindquist coordinates\cite{Johannsen:2010ru,Li:2022eue}
\begin{equation}
\begin{aligned}
r_i&=\sqrt{\alpha_0^2+\beta_0^2+d^2},\\
\theta_i&=\arccos{\frac{\beta_0\sin \theta_0+d\cos \theta_0}{r_i}},\\
\phi_i&=\arctan{\frac{\alpha_0}{d\sin \theta_0-\beta_0\cos \theta_0}},
\end{aligned}
\end{equation}
where $\theta_0$ is the inclination angle of the  photon as shown in Fig.\ref{obsfig}. The photons that can be observed on the image plane are those with momentum perpendicular to the image plane and against the line of sight, the components of the photon momentum $\vec{k}_0$ are given by
\begin{equation}
\begin{aligned}
k_{r}&=\frac{d}{r_i}k_0,\\
k_{\rm \theta}&=\frac{-\cos \theta_0+(\beta_0\sin \theta_0 + d\cos \theta_0)\frac{d}{r_i^2}}{\sqrt{ \alpha_0^2 + \left(d\sin \theta_0 - y'\cos \theta_0\right)^2}}k_0,\\
k_{\rm \phi}&=\frac{-\alpha_0\sin \theta_0}{\alpha_0^2+ (d\sin \theta_0 - \beta_0\cos \theta_0)^2}k_0.\nn
\end{aligned}\label{initialmoment}
\end{equation}

The radiation flux of accretion disk is closely related to circular geodesic motion of baryons on the equatorial plane. Obviously there are two conserved quantities for the baryons on circular orbits, energy and angular momentum. With the external spacetime metric, energy and angular momentum can be expressed as $p_t=-E=g_{tt}\dot{t}+g_{t\phi}\dot{\phi}$, $p_{\phi}=L=g_{t\phi}\dot{t}+g_{\phi\phi}\dot{\phi}$, with which the geodesic motion on the equatorial plane can be expressed as
\begin{align}
\frac{dt}{d\tau } &=\frac{E g_{\phi \phi }+L g_{t\phi}}{g_{t\phi }^{2}-g_{tt}g_{\phi \phi }}\,, \label{tdotmassive} \\
\frac{d\phi }{d\tau } &=-\frac{E g_{t\phi}+L g_{tt}}{g_{t\phi }^{2}-g_{tt}g_{\phi \phi }}\,, \label{phidotmassive}\\
g_{rr}\left( \frac{dr}{d\tau }\right) ^{2} &=-1+\frac{E^{2}g_{\phi \phi }+2ELg_{t\phi }+L^{2}g_{tt}}{g_{t\phi }^{2}-g_{tt}g_{\phi \phi }}\,.\label{rdotmassive}
\end{align}
In the above derivation we use the normalization condition $p_\alpha p^\alpha=m_0^2$ and set the mass of the massive particle $m_0$ to be 1. Note that $\dot{\theta}=0$, there is no particle motion in the direction of $\theta$. Eqs.(\ref{tdotmassive}) and (\ref{phidotmassive}) allow us to re-express energy and angular momentum  as
\begin{eqnarray}
E &=&\frac{g_{tt}-g_{t\phi}\Omega}{\sqrt{g_{tt}-2g_{t\phi}\Omega - g_{\phi\phi}\Omega^{2}}}\,, \\
L &=&\frac{g_{t\phi}+g_{\phi\phi}\Omega}{\sqrt{g_{tt} - 2g_{t\phi}\Omega - g_{\phi\phi}\Omega^{2}}}\,,
\end{eqnarray}
where $\Omega=d\phi/dt$ is the angular velocity of the massive particle on the equatorial plane. The circular equatorial orbits require $\dot{r}=\ddot{r}=0$, this condition along with Eq.(\ref{rdotmassive}) give rise to the circular orbits condition $g_{tt,r}+2g_{t\phi,r}\Omega+g_{\phi\phi ,r}\Omega^2=0$, from which one obtains the angular velocity
\begin{align}
\Omega =\frac{d\phi}{dt}=\frac{-g_{t\phi ,r}\pm\sqrt{(g_{t\phi,r})^{2}-g_{tt,r}g_{\phi \phi ,r}}}{g_{\phi \phi ,r}}\,.
\end{align}
From Eq.(\ref{rdotmassive}) one may introduce an effective potential
\begin{equation}
V_{\textmd{\scriptsize{eff}}}(r) = -1 + \frac{E^{2}g_{\phi\phi} + 2E L g_{t\phi} + L^{2}g_{tt}}{g_{t\phi}^{2} - g_{tt}g_{\phi\phi}}\,.
\end{equation}%
The position of the innermost stable circular orbit (ISCO) is determined by the effective potential through the conditions
\begin{align}
V_{\textmd{\scriptsize{eff}}}(r)=V_{\textmd{\scriptsize{eff}}}'(r)=V_{\textmd{\scriptsize{eff}}}''(r)=0\,.
\end{align}

In Ref.\cite{Page:1974he}, Page and Thorne considered the geometrically thin accretion disk, where the matter content is distributed on the equatorial plane. Under some reasonable assumptions and by application of the conservation laws of energy, angular momentum and rest mass, the radiation flux can be expressed as
\begin{align}
F(r)=-\frac{\dot{M}}{4\pi\sqrt{-g}}\frac{\Omega_{,r}}{\left(E-\Omega L\right)^2}\int_{r_{ms}}^r\left(E-\Omega L\right)L_{,r}dr\,,\label{flux}
\end{align}
which is the time-averaged flux of radiant energy flowing out of the face of accretion disk. Here $\dot{M}$ is the accretion rate, $r_{ms}$ is the radius of the ISCO or marginally stable circular geodesic orbit. We know the effect of redshift should be taken into account for the observed radiation intensity, the redshift factor is defined as $g=p_\mu u_o^\mu/p_\nu u_e^\nu$, where $p_\mu$ is the conjugate momentum of photon, $u_o^\mu$ is the four velocity of observer, and $u_e^\mu$ is the four velocity of emitting source.

To obtain shadow of Johannsen BH we need to trace the trajectories of photons that reach the observer. Since most of the photons would not reach the observer, direct method of tracing photons from the light source would waste lots of computation resources. Thus we use the open-source code Gyoto  which traces the light rays backwards from the image plane of observer to the plane of accretion disk\cite{Vincent:2011wz}. The trajectory integrations are performed according to Eqs.(\ref{eomphon}), which are solved numerically by the fourth order Runge-Kutta algorithm with an adaptive step. The initial momentum of the photon on the image plane is given by Eq.(\ref{initialmoment}). We modify the code to be applicable to Johannsen BH. Since  Johannsen BH possesses 4 deviation parameters, we would like to study the effects of the parameters on the accretion disk image one by one, like what we done in studying shadow boundary when examining one parameter we set other three parameters to be 0. We assume the image plane to be at large distance from the source BH, in the code we set the distance $d=100M$. We take two inclination angles $\theta_0=15^{\circ}$ and $\theta_0=60^{\circ}$ to show the accretion disk images on the image plane.

\begin{figure*}
\begin{center}
\includegraphics[width=0.45 \textwidth]{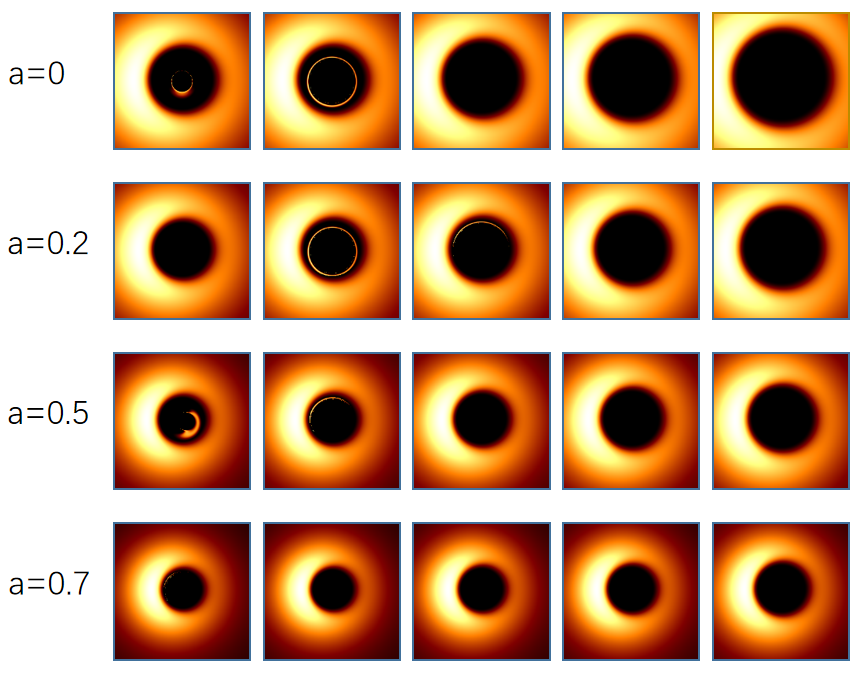}
\includegraphics[width=0.45 \textwidth]{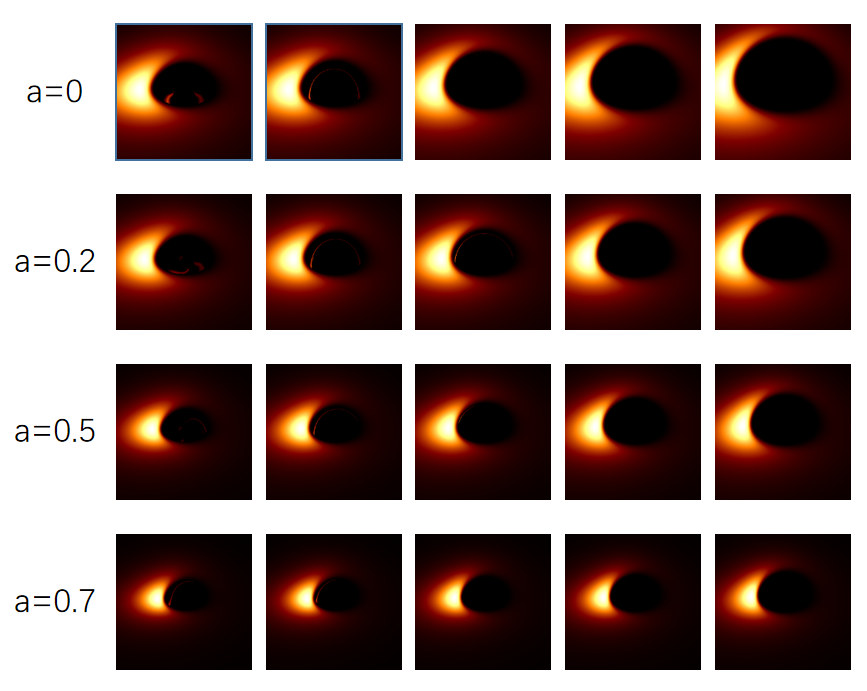}
\end{center}
\caption{Accretion disk images of Johannsen BH with different values of $\alpha_{13}$, $a$ and $\theta_0$. The left panel correspond to $\theta_0=15^{\circ}$, the right one correspond to $\theta_0=60^{\circ}$.  For the row $a=0$, from left to right $\alpha_{13}=-7, 0, 50, 100$ and $200$ respectively; for the row $a=0.2$, $\alpha_{13}=-7.5, 0, 10, 50$, $ 100$; for the row $a=0.5$, $\alpha_{13}=-6, 0, 10, 30$, $50$; and for the row $a=0.7$, $\alpha_{13}=-7.5, 0, 5, 10$, $20$. As comparision, the second columns on both panels correspond to the images of Kerr BH with $\alpha_{13}=0$. Masses of the black holes have been set to be $M=1$.}\label{accretiona13}
\end{figure*}

\begin{figure*}
\begin{center}
\includegraphics[width=0.45 \textwidth]{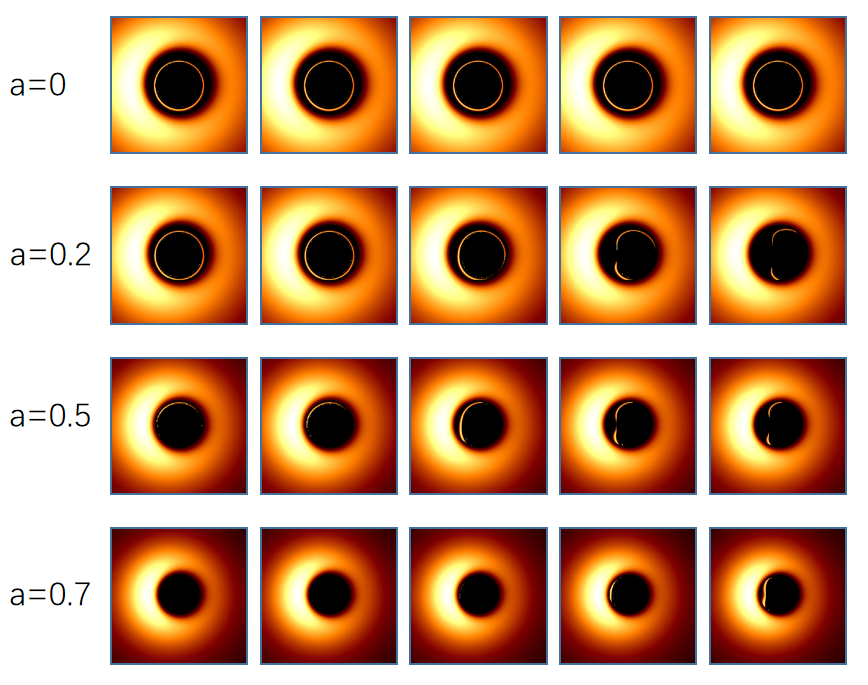}
\includegraphics[width=0.45 \textwidth]{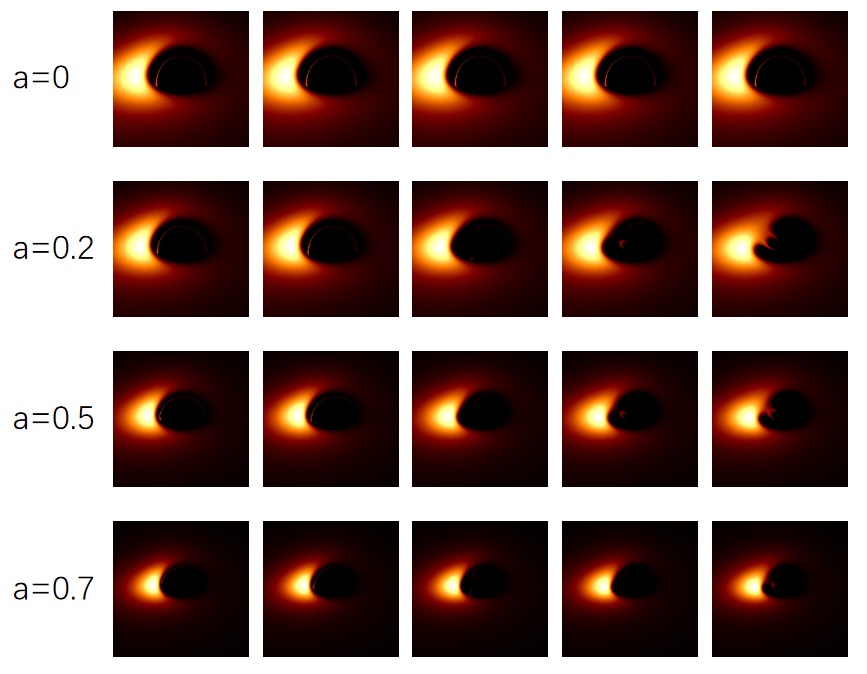}
\end{center}
\caption{Accretion disk images of Johannsen BH with different values of $\alpha_{22}$, $a$ and $\theta_0$. The left panel correspond to $\theta_0=15^{\circ}$, the right one correspond to $\theta_0=60^{\circ}$. For the row $a=0$, $\alpha_{22}=-3.9, 0, 100, 200$, $300$; for the row $a=0.2$, $\alpha_{22}=-3.8, 0, 50, 100$, $ 200$; for the row $a=0.5$, $\alpha_{22}=-3, 0, 20, 30$, $ 40$; and for the row $a=0.7$, $\alpha_{22}=-2.9, 0, 5, 10$, $ 15$. The second column on each panel corresponds to  Kerr BH with $\alpha_{22}=0$. We have set $M=1$.}\label{accretiona22}.
\end{figure*}

\begin{figure*}
\begin{center}
\includegraphics[width=0.45 \textwidth]{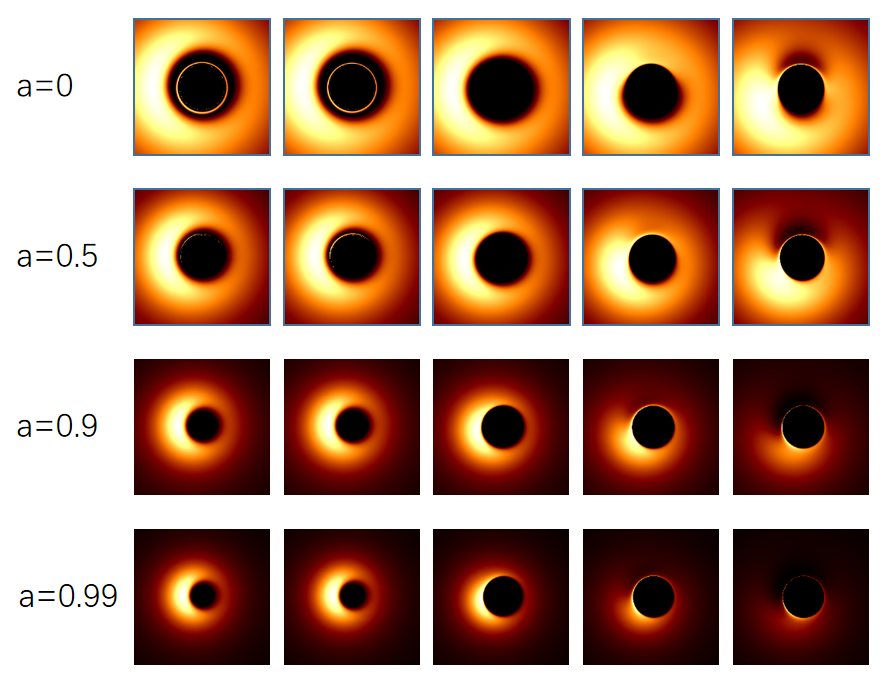}
\includegraphics[width=0.45 \textwidth]{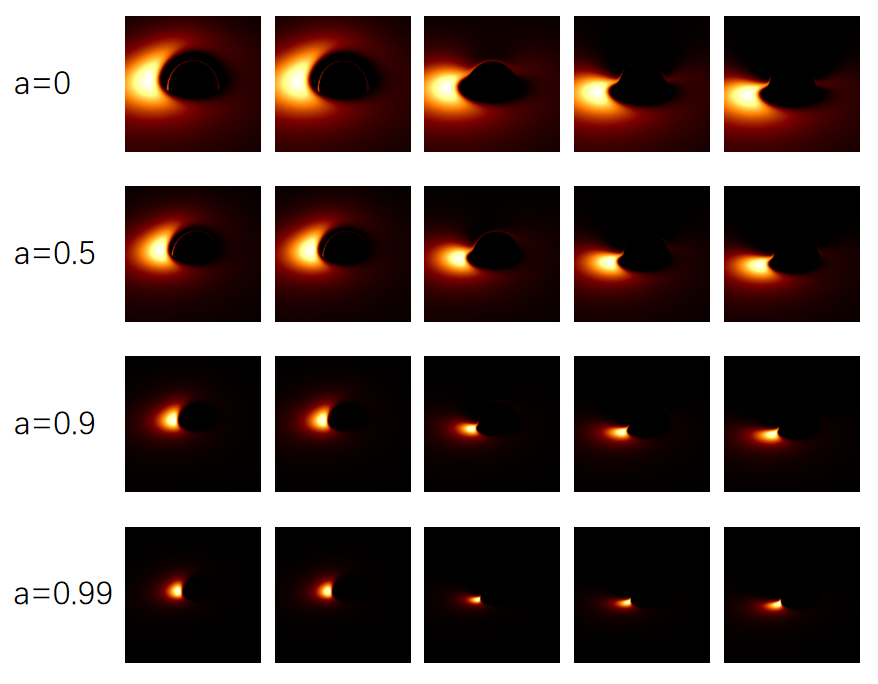}
\end{center}
\caption{Accretion disk images of Johannsen BH with different values of $\alpha_{52}$, $a$ and $\theta_0$. The left panel correspond to $\theta_0=15^{\circ}$, the right one correspond to $\theta_0=60^{\circ}$.  For the row $a=0$, $\alpha_{52}=-3.5, 0, 100, 200$, $300$; for the row $a=0.5$, $\alpha_{52}=-3, 0, 100, 200, 300$; for the row $a=0.9$, $\alpha_{52}=-2, 0, 100, 200, 300$; and for $a=0.99$, $\alpha_{52}=-1, 0, 100, 200, 300$. The second column on each panel corresponds to Kerr BH with $\alpha_{52}=0$. We have set $M=1$.}\label{accretiona52}.
\end{figure*}

\begin{figure*}
\begin{center}
\includegraphics[width=0.45 \textwidth]{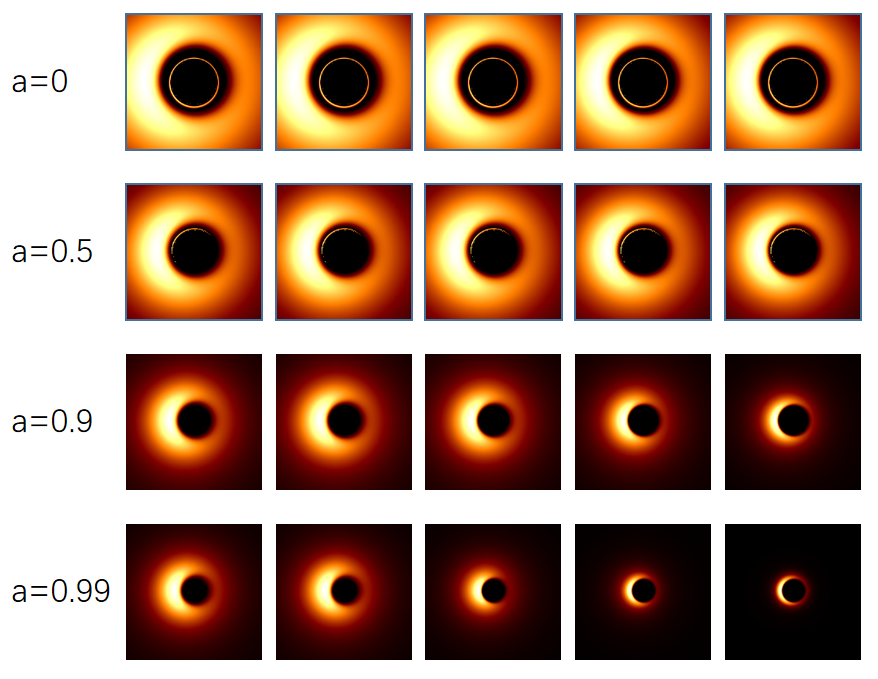}
\includegraphics[width=0.45 \textwidth]{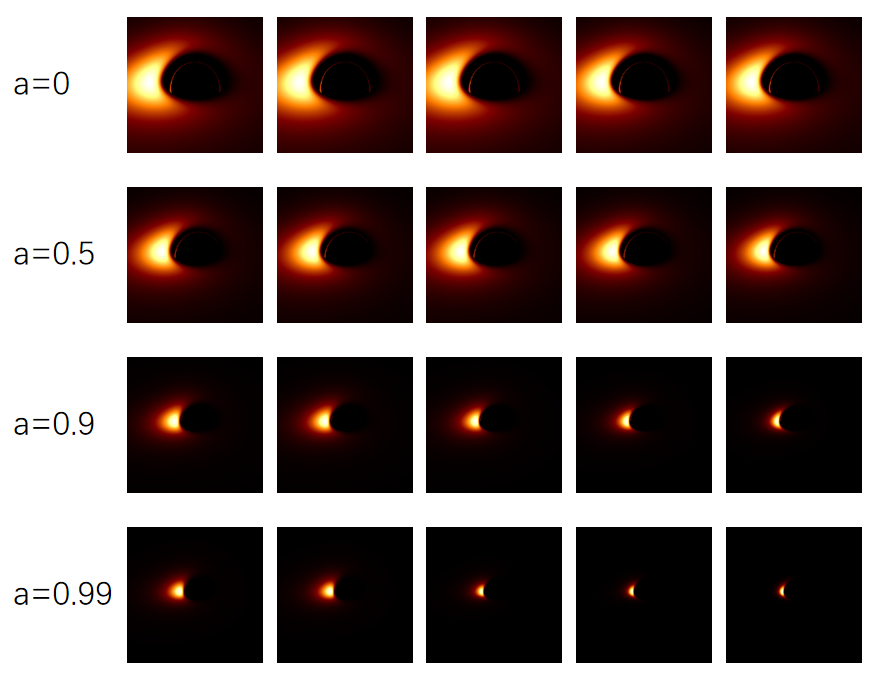}
\end{center}
\caption{Accretion disk images of Johannsen BH with different values of $\epsilon_3$, $a$ and $\theta_0$. The left panel correspond to $\theta_0=15^{\circ}$, the right one correspond to $\theta_0=60^{\circ}$. For the row $a=0$, $\epsilon_3=-7, 0, 25, 50, 100$; for the row $a=0.5$, $\epsilon_3=-6, 0, 10, 25, 50$; for the row $a=0.9$, $\epsilon_3=-2, 0, 10, 25, 50$; and for the row $a=0.99$, $\epsilon_3=-1, 0, 10, 25, 50$.  The second column on each panel corresponds to Kerr BH with $\epsilon_3=0$. We have set $M=1$.}\label{accretione3}.
\end{figure*}

Figs.\ref{accretiona13}-\ref{accretione3} exhibit the accretion disk images with different values of spin parameter $a$, inclination angle $\theta_0$ and deviation parameters $\alpha_{13}, \alpha_{22}, \alpha_{52}, \epsilon_{3}$. To facilitate the comparison between images with different BH parameters,  we put the images of Kerr BH in the second column in each figure. From Fig.\ref{accretiona13} one sees  that, $\alpha_{13}$ affects the sizes of the central dark regions of accretion disk images, the sizes of central dark regions increase with $\alpha_{13}$. For $\alpha_{13}>0$, the sizes of central dark region are larger than that of Kerr BH. The effects of $\alpha_{22}$ on accretion disk images are not so evident  as shown in Fig.\ref{accretiona22}. From Fig.\ref{accretiona52} one sees that, $\alpha_{52}$ affects both the brightness of the images and the sizes of central dark regions. The images become darker and darker as $\alpha_{52}$ increases, the positions of the regions with high luminosity vary with $\alpha_{52}$ as well. The central dark regions shrink with $\alpha_{52}$ for some values of $a$, but extend with $\alpha_{52}$ for other values of $a$. The shapes of central dark regions vary with $\alpha_{52}$ as well. Fig.\ref{accretione3} shows that, as $\epsilon_{3}$ increases the images become darker and the central dark regions shrink. All the Figs.\ref{accretiona13}-\ref{accretione3} exhibit that the central dark regions shrink with the spin parameter $a$. As we can see, the accretion disk images vary with BH parameters evidently, therefore the effects of BH parameters on BH shadow can be tested by  future  observations.

\section{Discussion and Conclusion\label{section5}}
Now let's discuss the results we obtained. Our calculations show that the shadow boundaries are affected by the deviation parameters $\alpha_{13}$ and $\alpha_{22}$. Though calculating the radii of shadows and comparing with the observations of M87* and SgrA*, we constrained the deviation parameters with BH image data motivated to test no-hair theorem. Our results showed that current observations cannot constrain the deviation parameters to be pure positive or pure negative, but constrain the deviation parameters to lie in a neighborhood of zero, which implies the current observations on BH image can't confirm the existence of BH hair. Our constraints on deviation parameters can be used to cross check with other constraints by other observations\cite{Tripathi:2018lhx,Tripathi:2020dni}.

We plot accretion disk image with ray-tracing method. The figures in our paper show that accretion disk images of Johannsen BH may differ from that of Kerr BH significantly.  For $\alpha_{13}<0$ the size of central dark region is smaller than that of Kerr BH. As $\alpha_{13}$ increases, the central dark region extends. For $\alpha_{13}>0$ the size of central dark region is larger than that of Kerr BH.  The effects of $\alpha_{22}$ are not very evident.
The deviation parameter $\alpha_{52}$ affects the brightness of the images. As $\alpha_{52}$ increases, the images become darker compared to that of Kerr BH, the central dark regions extend for some values of $a$ and shrink for other values of $a$. Moreover, the position of the shadow's bright region changes with $\alpha_{52}$.
Likewise, the deviation parameter $\epsilon_{3}$ affects the brightness of the images as well. As $\epsilon_{3}$ increases, the images become darker and the sizes of the central dark region become smaller compared to that of Kerr BH. The larger the value of the spin parameter $a$, the more evident the phenomenon is. When $a$ approaches 1, the images become very dark for large values of $\epsilon_{3}$ and $\alpha_{52}$. These properties can be used to test no-hair theorem with future observations.

In this paper, we investigate the shadow of a general rotating BH, that is, the Johannsen BH.  Johannsen BH is a parameterized generalization of Kerr BH which admits three constants of motion, i.e., mass, angular momentum and Carter constant.  The BH is characterized by mass, rotating parameters and  four deviation parameters $\alpha_{13}$, $\alpha_{22}$, $\alpha_{52}$  and $\epsilon_{3}$ to leading truncation order. Johannsen BH is not solution of any particular gravity theory, but it can be mapped to several known BH solutions of
 familiar theories of modified gravity. The universal features of Johannsen BH make it is very suitable for testing no-hair theorem in a model-independent way.

We plot the shadow boundaries on the plane of celestial coordinates. Among the four deviation parameters, only $\alpha_{13}$  and  $\alpha_{22}$ get into the expression of $R(r)$, $\alpha_{52}$  and $\epsilon_{3}$ don't appear in the expression of $R(r)$, thus the shadow boundaries are only affected by $\alpha_{13}$ and $\alpha_{22}$. The sizes of shadow boundaries grow with both $\alpha_{13}$ and $\alpha_{22}$ as can be seen from Fig.\ref{ShadowBoundary} and Fig.\ref{ShadowBoundary2}. The shadow boundaries are more deformed by larger rotating parameter $a$ and larger inclination angle $\theta_0$. The shadow boundaries are  deformed more greatly  by larger $\alpha_{22}$, but are deformed more greatly by smaller $\alpha_{13}$.

Since the two deviation parameters, $\alpha_{13}$  and  $\alpha_{22}$, affect the shadow boundary, they will also affect the shadow radius. Thus the observations on shadow radii can be used to constrain the deviation parameters. We use the observation results of M87* and SgrA* by EHT collaboration to constrain these parameters. Within the theoretically allowed ranges of the deviation parameters, the theoretical shadow radii varying with $\alpha_{13}$ may exceed both the upper and the lower observation bounds, which means too large and too small values of $\alpha_{13}$ are all disfavored. The shadow radii varying with $\alpha_{22}$ may exceed the upper observation bound while the lower observation bound are not exceeded, thus large values of $\alpha_{22}$ are disfavored. Therefore the deviation parameters are constrained in finite ranges with the observation results. There exist intersection points for the theoretical curves, on the left of the intersection points shadow radii are smaller for larger rotating parameters, while on the right of the intersection points the shadow radii are larger for larger rotating parameters.

Accretion disk is an important light source of the shadow, the accreted matter surrounding the central compact object emit photons which may reach the image plane of remote observer. Supposing the accretion disk to be a Page-Thorne thin disk, we draw the images of accretion disk through ray-tracing method. We study the effects of the BH parameters and learn that, the sizes of the central dark regions of the accretion disk images grow with $\alpha_{13}$, while the central dark regions are not affected evidently by $\alpha_{22}$. The shadows become darker as $\alpha_{52}$ or $\epsilon_{3}$ increases, and the position of the region with high luminosity changes with $\alpha_{52}$. Moreover, for fixed deviation parameters, the central dark regions shrink with increase of the rotating parameter $a$. Since the BH shadows are affected evidently by BH parameters, the upcoming observations of BH shadow can be used to constrain modified gravities in a unified way.

\section*{Acknowledgment}
We would like to thank N.Yang for useful discussions. This work is supported by the National Natural Science Foundation of China Grants Nos.11922303, 12275106, 12235019 and 12173071, and by Natural Science Foundation of Shandong Province No.ZR2023MA014, and The National Key R$\&$D Program of China (no. 2021YFC2203002). X. Fan is also supported by the Fundamental Research Funds for the Central Universitiesv(2042022kf1182). W. Han is also supported by CAS Project for Young Scientists in Basic Research YSBR-006.

\providecommand{\href}[2]{#2}\begingroup
\footnotesize\itemsep=0pt
\providecommand{\eprint}[2][]{\href{http://arxiv.org/abs/#2}{arXiv:#2}}

\end{document}